\documentclass[runningheads]{llncs}
\usepackage[table]{xcolor}

\usepackage[T1]{fontenc}
\usepackage{booktabs}
\usepackage{longtable}
\usepackage{array}
\usepackage{pdflscape}
\usepackage{tcolorbox}
\usepackage{here}
\usepackage{tabularx}

\usepackage{graphicx}
\newcommand{\rqanswer}[1]{%
\vspace{0.75em}
\noindent\raisebox{0.5em}{%
\begin{minipage}{\textwidth}
\begin{tcolorbox}[colback=gray!8,colframe=black,title=Answer]
#1
\end{tcolorbox}
\end{minipage}%
}%
\vspace{0.25em}
}

\begin{document}
\title{(How) Do Large Language Models Understand High-Level Message Sequence Charts?}
\titlerunning{Do LLMs Understand HMSCs?}
\author{Mohammad Reza Mousavi\inst{1}\orcidID{0000-0002-4869-6794}}
\authorrunning{M.R. Mousavi}
\institute{Department of Informatics, King's College London \\
\email{mohammad.mousavi@kcl.ac.uk}}
\maketitle              % typeset the header of the contribution
\begin{abstract}
Large Language Models (LLMs) are being employed widely  to automate tasks across the software development life-cycle. It is, however, unclear whether these tasks are performed consistently with respect to the semantics of the artefacts being handled. This question is particularly under-researched concerning architectural 
design specification. 

In this paper, we address this question for High-level Message Sequence Charts (HMSCs). These are visual models with a rigorous formal semantics that have been used for various purposes, including as a foundation for Sequence Diagrams in the Unified Modelling Language (UML). We examine whether LLMs ``understand'' the semantics of HMSCs by examining three LLMs (Gemini-3, GPT-5.4, and Qwen-3.6) on how they perform 129 semantic tasks ranging from querying basic semantic constructs in HMSCs (i.e., events and their ordering) to semantic-preserving abstractions and compositions, and calculating the set of traces and trace-equivalent labelled transition systems. 
The results show that 
LLMs only have a modest understanding 
of the formal semantics of HMSCs 
($\sim$52\% overall accuracy), 
with great variability across different
semantic concepts: 
while LLMs seem to understand the basic semantic concepts of MSCs ($\sim$88\% accuracy), they struggle with semantic reasoning in tasks involving abstraction and composition ($\sim$36\% accuracy) and traces and LTSs ($\sim$42\% accuracy). In particular, all three LLMs struggle with the notions of co-region and explicit causal dependencies and never employed them in semantic-preserving transformations. 

\end{abstract}

\begin{minipage}[t]{0.48\textwidth}
\end{minipage}
\hfill
\begin{minipage}[t]{0.48\textwidth}
This paper is dedicated to \\ Sjouke Mauw and Michel Reniers.\\ To Sjouke Mauw, \\ on the occasion of his retirement. \\ To Sjouke Mauw and Michel Reniers, \\ for their contributions to \\ the formal semantics of \\ Message Sequence Charts;  and   \\
    To Michel Reniers, \\  for teaching me formal semantics....
    
\end{minipage}

\begin{flushright}

\end{flushright} 

\section{Introduction}

\subsection{Motivation}
High-level Message Sequence Charts (HMSCs) \cite{MSC1993} are among the pioneering models for software and system specification, which are furnished with a formal semantics. They have been widely used in practice, standardised, and have also been integrated into the Unified Modelling Language \cite{BoochUML1999} in the form of sequence diagrams.

\subsection{Problem Definition}
\emph{Understanding} architectural specification is among the challenging usecases for Large Language Models (LLMs) \cite{GroplerKJDSDTNMTVALZZ25}.
By \emph{``understanding''} we do not mean to anthropomorphise LLMs \cite{Salles02042020}; rather, we are referring to their observable behaviour: demonstrating outputs that are consistent with a sound model of the semantics of HMSCs. 
It is unclear whether LLMs respect the semantics of architectural 
specification languages, even with respect to their most fundamental semantics construct. 

\subsection{Approach} 
In this study, our aim was to quantify the LLMs' understanding of HMSCs.
To this end, we identify the key semantic concepts in the formal semantics of HMSCs; formulate and carry out tasks that examine their semantic understanding;  and analyse the performance of LLMs and provide a quantification of the understanding of various semantic concepts.

\paragraph{Key semantic concepts.} The formal semantic concepts are drawn from the formalisation of HMSCs, based on the ITU Recommendation Z.120 Annex B: Algebraic Semantics of Message Sequence Charts \cite{MSC1993}.
The identified semantics concepts are: causal ordering of events, abstraction, composition, traces, and (trace-equivalent) labelled transition systems. We expand on each of these concepts  and identify patterns that exemplify them in the subsequent explanation, generation, and manipulation tasks.

\subsection{Results}
Our results indicate that the three LLMs evaluated in our paper, all have a modest 
understanding 
of the formal semantics of HMSCs 
($\sim$52\% overall accuracy), 
with great variability across different
semantic concepts: 
they all have a 
precise model of the basic semantics constructs of  (H)MSCs such as  their comprising events and event orderings  ($\sim$88\% accuracy). However, they struggle with abstraction, composition ($\sim$36\% accuracy), and higher-level reasoning about behavioural semantics ($\sim$42\% accuracy). They all seem to lack a sound notion of co-region (a region with no ordering of events) and explicit causal orders among events. For example,  they never use co-regions when composing (H)MSCs horizontally (in parallel), leading to introducing fictitious orders in such compositions. Also, when abstracting from events, they cannot reinstate the ordering implied by the abstracted events, leading to fewer orders than the original (H)MSC.  Another interesting observation is that in the few cases of successful higher-level reasoning, they generate Python code for the corresponding reasoning task and execute it to obtain a sound result. This indicates that potentially reliable results can be obtained by interfacing LLMs with other tools and encoding the reasoning tasks into the input of such external tools.  The details of experiments, results, and analysis are publicly available through the replication package: \url{https://smrmousavi.github.io/sjouke-fest/}. 

\subsection{Organisation}
The remainder of this paper is organised as follows. 
We review the related work to position our work with respect to the literature on understanding code and architectural diagrams in the context of Large Language Models in Section \ref{sec:related}. 
Then we recall some basic concepts regarding the syntax and semantics of (H)MSCs in Section \ref{sec:hmsc}.
In the same section, we identify the key semantic concepts that we would like to examine in our experiments. 
This leads to a formulation of our research questions and the design of experiments to answer them in Section \ref{sec:methodology}.
We report the results of our experiments and answer the formulated research questions in Section \ref{sec:results}. 
Finally, we reflect on these answers and on the threats to the validity and the generalisability of the results in Section \ref{sec:reflection}.

\section{Related Work}\label{sec:related} 
In this section, we review the related lines work on measuring semantic understanding in LLMs for code and for architectural specifications. 

\paragraph{Understanding Program Semantics.} 
Although understanding code semantics may deal with more fine-grained tasks than the architectural tasks studied in this paper, many of the semantic concepts  examined in this paper are still relevant at code-level (operational) semantics, in particular for reasoning about concurrency. 

Ma et al.\ \cite{ma2024} analyse the LLM capabilities in code understanding at three levels: syntax, static semantics, and dynamic semantics. For understanding dynamic semantics, they report an accuracy of ca.\ 60\% using zero-shot (and slightly lower using few-shot) prompting.  Laneve et al. \cite{Laneve2026} analyse program semantic understanding among seven different LLMs. They do so by generating tasks about program  equivalence and analysing the correctness of the outcome. They report a mean accuracy of  59\% in zero-shot prompting and 71\% after providing contextual information about the perturbation in the programs.  They conclude that there is a significant gap in the semantic reasoning capabilities of the analysed LLMs.  
Similarly, Wei et al.\ \cite{wei2025} analyse the capabilities of 19 LLMs to reason about the programme equivalence and report a mean precision of 64\%. 
Hooda et al.\ \cite{Hooda2024} use counter-factual analysis to analyse LLM program understanding: they use four different Program Concept Predicates to characterise four classes of semantic equivalence in programs. They use these concepts to evaluate the performance of LLMs in code completion and program repair tasks. Their analysis shows significant gaps in understanding program semantics that influence downstream task precision by up to 33\%. 
All of these approaches address the issue of semantic understanding, and point out significant gaps in understanding code semantics. Our methodology differs from all of these approaches, since they use  a typical downstream task such as equivalence checking or code completion to analyse semantic understanding. However, in our approach, we directly test the semantic concepts such as ordering and traces in our queries, which give us a more direct access to semantic understanding metrics. 

Nikiema et al.\ \cite{nikiema2025} used obfuscation to measure semantic understanding; 
they used the accuracy of generating explanation and 
de-obfuscation as proxies for understanding and 
showed that code obfuscation leads to significant decline in the accuracy of both tasks.

In another line of enquiry, Spiess et al.\ \cite{spiess2026} study the robustness of code semantic understanding. They use input perturbations to check the robustness of output prediction tasks as a proxy for understanding dynamic semantics. Their results, in line with all the previous line of research, show lack of robustness in the understanding of dynamic semantics. 

In a different line of work, North et al.\ \cite{North2025} analyse whether there is a common semantic understanding in LLMs across different programming languages. Their initial results suggest that such a common semantic understanding does exist.  
While this problem is orthogonal to the problem studied in this paper, further insights about the common semantic understanding across different languages can prove essential; it may enable  exploring the boundaries of semantic understanding in LLMs and extending and augmenting it. 

Table \ref{tab:code-semantics-comparison} provides 
an overview of the comparison of related work in code semantic understanding; although our paper is not directly about programming semantics, our approach and results can be useful in this domain as identified in the last row. In particular, our methodology in formulating semantic tasks that directly examine the basic semantic concepts is complementary to all of the reviewed pieces of literature in Table \ref{tab:code-semantics-comparison} and can be employed to devise novel studies that go beyond the available research. 

\begin{table}[t]
\centering
\scriptsize
\caption{Comparison with work on code semantic understanding.}
\label{tab:code-semantics-comparison}
\begin{tabular}{p{0.3\textwidth} p{0.3\textwidth} p{0.4\textwidth} }
\toprule
\textbf{Work} & \textbf{Type of study} & \textbf{Metrics used} \\
\midrule
Ma et al.\ \cite{ma2024} & Code syntax/semantics & Accuracy \\
Laneve et al.\ \cite{Laneve2026} & Program equivalence & Accuracy \\
Wei et al.\ \cite{wei2025} & Program equivalence & Precision \\
Hooda et al.\ \cite{Hooda2024} & Counterfactual predicates & Downstream precision \\
Nikiema et al.\ \cite{nikiema2025} & Code obfuscation & Task accuracy  \\
Spiess et al.\ \cite{spiess2026} & Execution perturbations & Robustness/accuracy   \\
North et al.\ \cite{North2025} & Cross-language semantics & Comparative performance\\
\rowcolor{gray!15}
\textbf{Our paper} & \textbf{HMSC semantic benchmark} & \textbf{Stratified accuracy}  \\
\bottomrule
\end{tabular}
\end{table}

\paragraph{Architectural Understanding.} Schmid et al. \cite{schmid2025} perform a systematic review of the use of LLMs in architectural tasks in software engineering. The results indicate that there is very limited research on the suitability of LLMs for architectural understanding. Below we review some of the recent literature addressing this research gap. 

Soliman et al. \cite{Soliman2025b,Soliman2025a} study the availability and accuracy of architectural knowledge in LLMs. They evaluate three types of architectural questions and report a precision of below 20\% and a recall of 20\%-40\% among seven LLMs. Corroborating this report,  Zhou et al.\ \cite{Zhou2025} report a precision of ca. 27\% and recall of ca. 66\% regarding explanations of design rationale. Elberzhager et al.\ \cite{Elberzhager2026} use case studies to evaluate the consistency and accuracy of  LLM-supported evaluation of architecture documents. The provided analysis is brief and is confined to two case studies, but they suggest high consistency and low precision. 
Gurerra and Ernst \cite{GuerraErnst2025} analyse the performance of LLMs with respect to  architectural tasks; they categorise the tasks into five levels: remembering, understanding, applying, analysing, evaluating, and creating. They use a case study for their experiment and they report surprisingly high precision scores (called grades) across all tasks, with slightly lower scores on  the initial (lower-level) tasks. We expect these high scores are motivated by the generic nature of the analysed case study and the fact that LLMs have potentially been trained on similar architectures. 

Ouyang et al.\ \cite{ouyang2026} study architectural understanding from a multi-modal perspective by evaluating Vision-Language Models on software architecture diagrams. They introduce SADU, a benchmark comprising 154 behavioural, structural, and entity--relationship diagrams with 2,431 question-answering tasks targeting counting and retrieval reasoning over architectural elements. Their evaluation of 11 state-of-the-art VLMs shows that diagram-based architectural understanding remains challenging: the best model reaches only 70.18\% accuracy, with performance degrading on more complex diagrams and on visually difficult relation structures such as overlapping or non-standard arrows. The current paper complements the study by Ouyang et al.\ \cite{ouyang2026} in exploring the boundaries of LLM semantic analysis capabilities.

Table \ref{tab:architecture-understanding-comparison}
summarises the comparison of the related work on architectural understanding. 
It appears from the existing literature that we lack a fundamental understanding LLMs capabilities in architectural understanding. In particular, we do not seem to have a precise characterisation of the boundaries of correct architectural reasoning in LLMs. The goal of the present paper is to take the first step in this direction for the specific case of HMSCs. 

\begin{table}[t]
\centering
\scriptsize
\caption{Comparison with work on architectural understanding.}
\label{tab:architecture-understanding-comparison}
\begin{tabular}{p{0.18\textwidth} p{0.24\textwidth} p{0.20\textwidth} p{0.28\textwidth}}
\toprule
\textbf{Work} & \textbf{Type of study} & \textbf{Metrics used} & \textbf{Gap filled by our paper} \\
\midrule
Schmid et al.\ \cite{schmid2025} & Systematic review & Literature synthesis & No semantic benchmark \\
Soliman et al.\ \cite{Soliman2025b,Soliman2025a} & Architecture knowledge QA & Precision, recall & General knowledge only \\
Zhou et al.\ \cite{Zhou2025} & Design-rationale generation & Precision, recall & Rationale, not semantics \\
Elberzhager et al.\ \cite{Elberzhager2026} & Document case studies & Consistency, precision & Document-level evaluation \\
Guerra and Ernst & Architecture task taxonomy & Grades/precision & Broad case-study tasks \\
Ouyang et al.\ \cite{ouyang2026} & Architecture diagram VQA & Accuracy & Visual grounding only \\
\rowcolor{gray!15}
\textbf{Our paper} & \textbf{HMSC semantic benchmark} & \textbf{Stratified accuracy} & \textbf{Direct formal semantics} \\
\bottomrule
\end{tabular}
\end{table}

\section{High-Level Message Sequence Charts}\label{sec:hmsc}

\begin{figure}
\begin{center}

\includegraphics[scale=0.4]{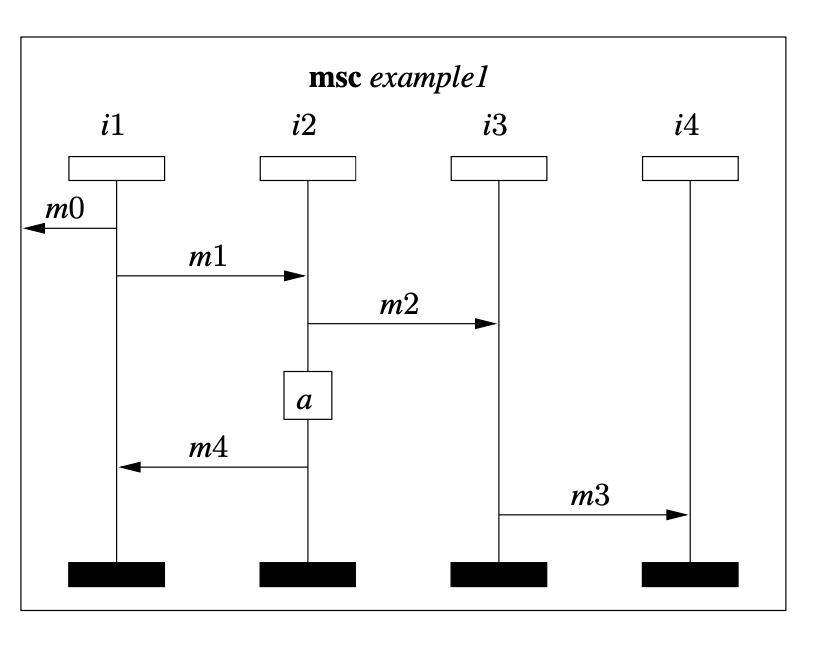}
\caption{An MSC labelled ``example1'', with four instances, labelled ``i1'' to ``i4'', five messages labelled ``m0'' to ``m4'', and an internal action labelled ``a''.}\label{fig:example1MSC}
\end{center} 

\end{figure}

\subsection{Syntax}
A message sequence chart (MSC) is given by a labelled frame (a bounding box) comprising a number of instances, represented by vertical lines labelled by the instance names. Intuitively, they can be thought of as objects in a software system, or nodes in a distributed system. They send and receive messages, denoted by labelled arrows across instances. They may also perform internal actions, denoted by squares placed on vertical lines. Message can, for example, represent method calls in  software or packet send and receive actions on a network.
In addition to messages sent and received between instances, they can also be sent and received to and from the environment denoted by arrows starting from and ending at the frame, respectively. 
 An example of an MSC is depicted in Figure \ref{fig:example1MSC}. 
This is the first subject system used in our experiments. Below we explain the syntactic elements used in this example. 

\begin{example}[MSC: Syntax]
Figure \ref{fig:example1MSC} depicts an MSC labelled ``example1''. 
It comprises four instances, labelled ``i1'' to ``i4''; messages ``m1'' to ``m4'' are sent and received between instances, while message ``m0'' is sent to the environment. Internal action ``a'' takes place within instance ``i2''.
Identifying the events in this example was the first task defined in our experiments. 
\end{example}

\begin{figure}
\begin{center}

\includegraphics[scale=0.3]{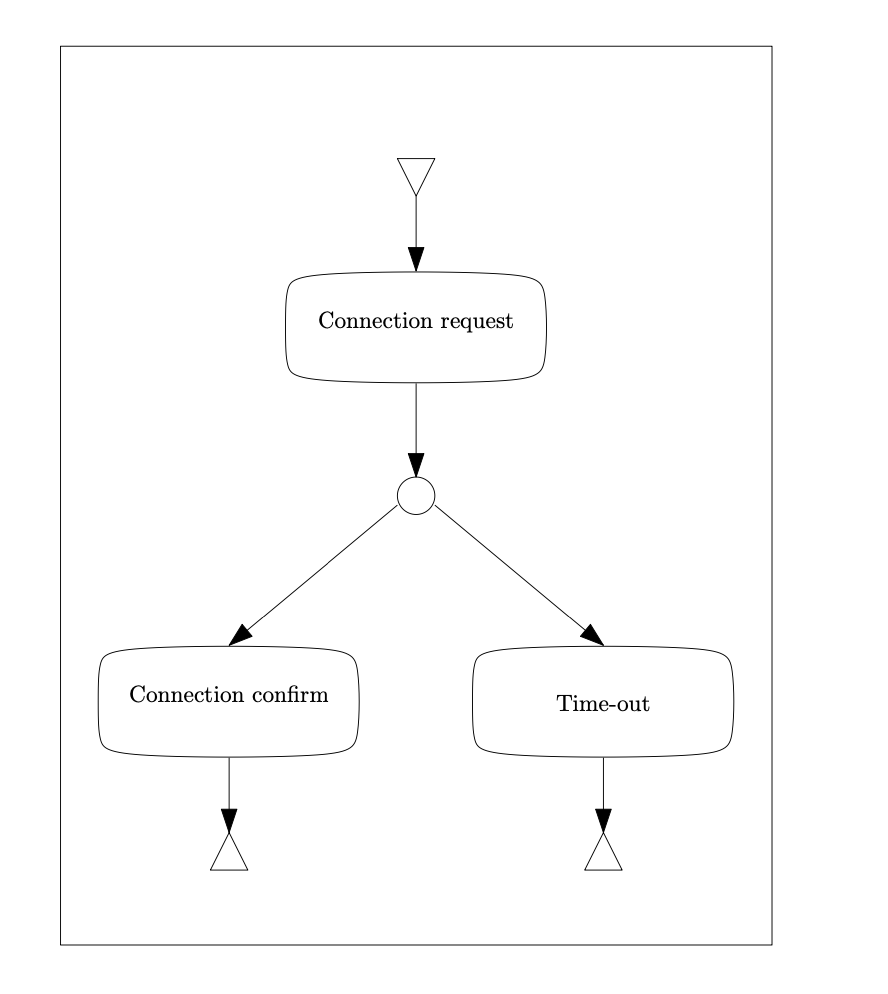}
\caption{An HMSC labelled comprising a vertical composition of ``Connection request'' with a choice between ``Connection confirm'' and ``Time-out''.}\label{fig:example7HMSC-part-i}
\end{center} 

\end{figure}

\begin{figure}
\begin{center}

\includegraphics[scale=0.5]{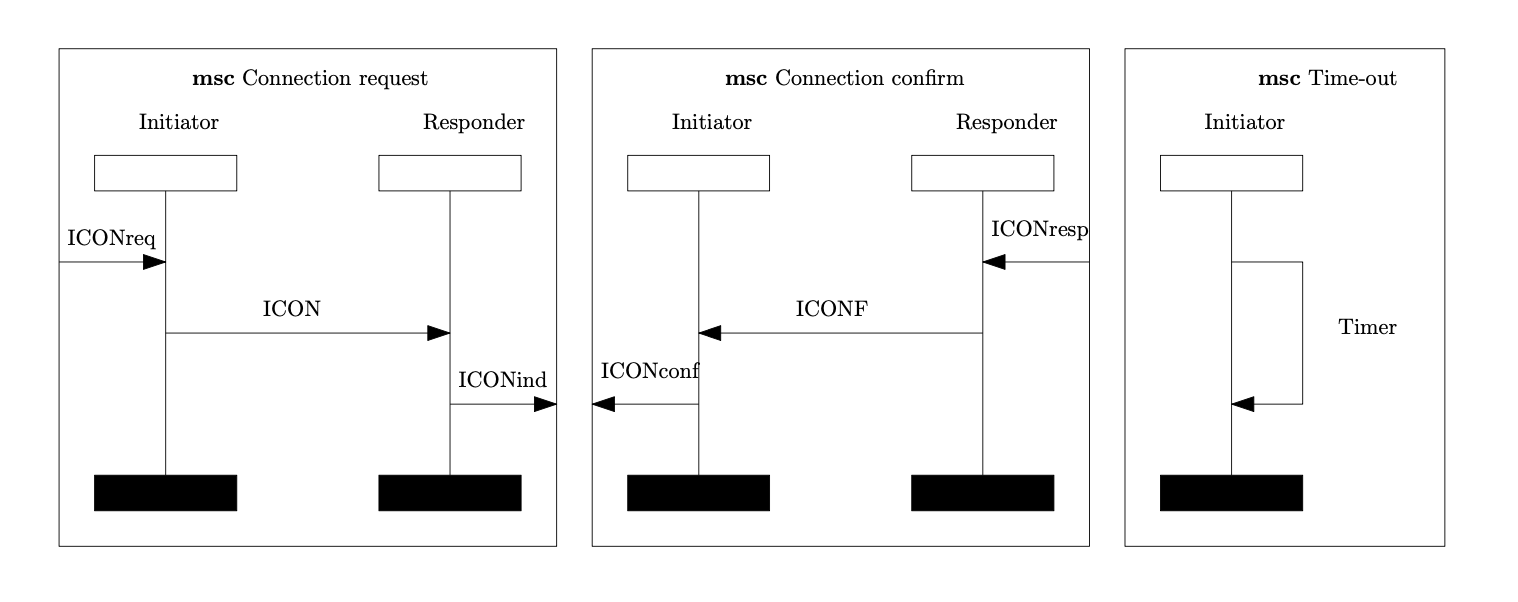}
\caption{The MSCs referenced in the HMSC of Figure \ref{fig:example7HMSC-part-i}.}\label{fig:example7HMSC-part-ii}
\end{center} 

\end{figure}

HMSCs are the result of composing MSCs using choice operator, depicted by a branch, vertically (sequentially), denoted by sequencing, and horizontally (in parallel) denoted by juxtaposing the MSCs.  
Choice and sequential composition can be combined to represent iteration (loop). 
Figures \ref{fig:example7HMSC-part-i} and \ref{fig:example7HMSC-part-ii} depict an HMSC and its components MSCs, respectively. 

\begin{example}[HMSC: Syntax]
Figure \ref{fig:example7HMSC-part-i} depicts an HMSC that starts off with the behaviour specified by the MSC labelled ``Connection request'', vertically (sequentially) composed by a choice between ``Connection confirm'' and ``Time-out''. The MSCs referenced in this HMSC are depicted in Figure \ref{fig:example7HMSC-part-ii}. 
\end{example}

\subsection{Semantics}

The identified key semantic concepts are listed below:
\begin{enumerate}
%\items Events: the building block of s
\item Causal ordering: the most important semantic concept in Basic Message Sequence Charts (BMSCs, the building blocks of HMSCs) is the causal ordering of events; they comprise the following types of orders:
\begin{enumerate}
    \item Send and receive order: a message cannot be received before it is sent; 
    \item Vertical order: events in the same instance and on the same vertical line are totally ordered from top to bottom;
    \item Transitive closure: causal ordering of events is a transitive relation.  
\end{enumerate}
\item Abstraction: an event can be abstracted from a BMSC, leading to an equivalent BMSC with fewer events but the same causal ordering among the remaining events; moreover, entities can be merged to create BMSCs with fewer entities. 

\item Composition: MSCs can be composed using various composition operators: horizontal composition semantic enforces that events 
in common components maintain the order specified in the individual
components, but are unordered between the two components.  
Vertical composition semantic defines that events 
in common components maintain the order specified in the individual
components, and additionally, those in the instance of the first component 
all happen before those in the common instance of the second component. 
Choice means that there are two sets of partial orders defined by each of the two components of the choice. 
The semantics of loops is defined in terms of a choice among the unrollings of the loop (eventually leading to choice and vertical composition). 

\end{enumerate}

Below we explain the semantic concepts relevant for the (H)MSCs depicted in Figures \ref{fig:example1MSC}, \ref{fig:example7HMSC-part-i}, and \ref{fig:example7HMSC-part-ii}. We start with illustrating the basic semantic concepts of MSCs. 

\begin{example}
Consider the MSC depicted in Figure \ref{fig:example1MSC}.
 Sending ``m0'' happens before sending ``m1'' because they are ordered by the vertical line of instance ``i1''. Sending ``m1'' happens before receiving ``m1'' due to the causal order between send and receive. Sending ``m1'' happens before receiving ``m3'' due to the causal order in the sends and receives for ``m1'', ''m2'', and ``m3'', the vertical order between receiving ``m1'' and sending ``m2'' and receiving ``m2'' and sending ``m3'',  and finally the transitive closure of all these orders. 
In fact, this is the second task given to the three LLMs to examine their semantic understanding. 
\end{example}

Next, we illustrate the semantics of HMSCs. 

\begin{example}
Consider the HMSC depicted in Figures \ref{fig:example7HMSC-part-i} and \ref{fig:example7HMSC-part-ii}.
The semantics of choice defines two partial orders for the HMSC: 
one resulting from vertical composition of 
``Connection request'' and ``Time-out'' and 
another from the vertical composition of  ``Connection request''  and ``Confirm''. 
Focussing on the former, this means that the partial order will be the union of the partial orders defined by the semantics of the two constituent MSCs as well as additional orders between receiving ``ICONreq'' and sending ``ICON'' on one hand, sending and receiving ``Timer'' on the other hand. 
\end{example}

The event orders specified above give rise to a set of traces, a totally-ordered sequence of events respecting the orders. Moreover, the semantics of the MSC can be presented by a labelled transition system capturing all the events and event ordering specified by the semantics (modulo trace equivalence).

\section{Methodology}
\label{sec:methodology}

This section describes the methodology used to evaluate the extent to which contemporary LLMs can understand and reason about the formal semantics of HMSCs. The study is organised around a controlled benchmark of subject models, a set of semantic reasoning and transformation tasks, and a comparative evaluation across multiple LLMs.

\subsection{Research Questions}
\label{subsec:research-questions}

The study is guided by the following research questions:

\begin{description}
    \item[RQ1:] To what extent can LLMs identify and reason about the basic semantic elements of MSCs?

    This question evaluates whether LLMs can recognise the foundational building blocks of the semantics  of MSCs, including events (send, receive and internal actions),  and causal ordering relations (induced by send and receive,  by the vertical order, and by the transitive closure of the former two).
    
    \item[RQ2:] How well do LLMs perform semantic transformations over MSCs?

    This question concerns operations that require more than having a notion of the basic semantic concepts. The evaluation includes transformations such as abstraction from events and merging of instances, where the resulting MSC must preserve the relevant behavioural semantics of the original chart.

    \item[RQ3:] Can LLMs reason about the behavioural semantics of MSCs?

    This question assesses whether LLMs can derive behavioural representations from MSCs. In particular, it considers whether models can identify the set of traces represented by an MSC and generate a labelled transition system with equivalent trace semantics.

    \item[RQ4:] How does the performance vary across LLMs model complexity and task complexity?

    This question investigates whether the accuracy varies across different LLMs and whether the performance of LLMs depends on the complexity of the subject HMSC and on the complexity of the task being performed. 

    \item[RQ5:] Which MSC concepts constitute recurring ``pain patterns'' for LLMs?

    This question aims to identify classes of semantic reasoning problems that systematically challenge LLMs. 
   
\end{description}

\subsection{Subject Systems}
\label{subsec:subject-systems}

The subject systems consist of nine MSCs and HMSCs taken from Software Specification lecture notes taught by Michel A. Reniers and the author through six years at TU/Eindoven. \footnote{They are all available publicly in the replication package: \url{https://smrmousavi.github.io/sjouke-fest/}.}  
These models were selected because they provide a controlled and pedagogically validated corpus of examples. 
Their use offers two main advantages. 
First, there is a high degree of confidence that the models are well formed. Second, solutions to the corresponding exercises are available, allowing the generated LLM responses to be compared against reference answers.

The benchmark covers both basic MSC- and HMSC semantic concepts. The basic concepts include events, send and receive actions, internal events, causal ordering, vertical ordering along instances, transitive closure, abstraction from events, merging of instances, and co-regions. The higher-level concepts include horizontal and vertical composition, HMSC notation, and composition operators.

A limitation of this choice of subject systems is that the models originate from lecture notes, and some LLMs may have been exposed during training to unofficial copies of these materials. This potential data contamination risk is therefore considered a threat to validity.

\subsection{Evaluated Models}
\label{subsec:evaluated-models}

The evaluation compares the performance of three contemporary LLMs:

\begin{itemize}
    \item Gemini-3-Pro Thinking (Copilot Interface);
    \item GPT-5.4  Think Deeper;
    \item Qwen-3.6-Plus Thinking.
\end{itemize}

These models were selected as representative examples of state-of-the-art generative AI systems. The purpose of comparing multiple LLMs is to determine whether the observed strengths and weaknesses are model-specific or whether they reflect broader limitations in current LLM-based reasoning about MSCs and HMSCs.

\subsection{Evaluation Tasks}
\label{subsec:evaluation-tasks}

The evaluation tasks are derived from the semantic concepts underlying MSCs and HMSCs. Each task is expressed as a natural-language prompt and accompanied by a diagrammatic MSC or HMSC input where appropriate.

\subsubsection{Event Identification}

The first class of tasks asks the model to identify the events present in a given MSC. These tasks evaluate whether the model can correctly recognise the basic syntactic and semantic units of an MSC.

\subsubsection{Causal Ordering}

The second class of tasks evaluates whether the model can determine ordering relations between events. Correct answers require the model to account for send--receive order, vertical ordering along instances, and the transitive closure of these relations.

\subsubsection{Abstraction from Events}

The third class of tasks concerns abstraction. In these tasks, the model is asked to generate or reason about an MSC in which a given message or event has been abstracted away. Such tasks test whether the model can preserve the correct observable behaviour after hiding or removing specified events. They also require the model to update causal dependencies appropriately after abstraction.
Also in the same category,  we ask the model to merge two instances in an MSC and produce the resulting chart. This evaluates whether the model can correctly combine lifelines while preserving the event ordering and communication dependencies induced by the original MSC.

\subsubsection{Trace Identification}

The fifth class of tasks requires the model to compute the set of traces represented by an MSC. These tasks assess the model's ability to reason about concurrency and interleaving. Since MSCs often denote partial orders rather than single total orders, identifying all possible traces requires systematic reasoning about the linear extensions of the underlying partial order.

\subsubsection{Generation of Labelled Transition Systems}

The final class of tasks asks the model to generate a labelled transition system with the same traces as the given MSC. These tasks require translation from a diagrammatic specification to an operational behavioural representation. They therefore evaluate whether the model can move beyond recognition of diagram elements and construct an equivalent formal model.

\subsection{Experimental Procedure}
\label{subsec:experimental-procedure}

The experimental procedure consists of the following stages.

\begin{enumerate} 

\item \emph{Benchmark construction.} 
A benchmark of nine MSC/HMSC subject models is constructed from Software Specification lecture notes. 

\item \emph{Prompt design.}
For each subject model, task-specific natural-language prompts are created. 

\item \emph{LLM execution.}
Each prompt is submitted to each of the three evaluated LLMs. The models produce textual, symbolic, or diagrammatic responses, depending on the nature of the task. All generated responses are recorded for subsequent comparison with reference solutions.

\item \emph{Reference-based assessment.}
The generated outputs are assessed manually against the available reference solutions. We had to  make numerous judgement calls and filter out syntactic variations that had no semantic significance (e.g., repetitions in sets). Prior to the assessment, we performed several rounds of validation 
(resulting in four rounds of correction and re-execution) to ensure the 
consistency among the tasks executed across models. The results of validation rounds are recorded in the replication package.

\item \emph{Aggregation of results.}
Correctness judgements are aggregated to obtain overall accuracy scores. The results are further stratified by LLM, model complexity and task complexity. This allows the evaluation to distinguish between performance on simple and complex MSCs, and between performance on basic and advanced reasoning tasks.

\item \emph{Error analysis and reflection.}
Finally, incorrect responses are analysed to identify recurring failure modes.

\end{enumerate}

The primary evaluation metric is accuracy, defined as the proportion of tasks for which the LLM produces a correct answer according to the reference solution.

\subsection{Pipeline Implementation}
\label{subsec:pipeline-implementation}

The evaluation pipeline implements the above procedure as a sequence of benchmark preparation, prompt execution, response collection, semantic assessment, and result aggregation steps. First, the MSC/HMSC diagrams and their corresponding reference solutions are collected from the lecture-note corpus. Second, for each diagram, a set of prompts is instantiated according to the task categories described above. Third, each prompt is submitted independently to each evaluated LLM. Fourth, the generated responses are compared with the reference solutions using task-specific correctness criteria. Finally, the resulting correctness judgements are aggregated into overall and stratified accuracy measures.

The pipeline is intentionally designed around semantic rather than purely syntactic correctness. For example, in abstraction tasks, a response is not judged solely by whether it removes the requested message, but by whether the resulting MSC preserves the correct observable behaviour. Similarly, in trace-generation tasks, a response is judged by whether it captures all and only the traces permitted by the causal structure of the original MSC.

\section{Results}\label{sec:results}

Prior to creating stratified accuracy results across LLMs, task types and HMSC model complexity, 
we performed a number of validation rounds. We do not present the results of our validation, but make them available through the replication package (both in raw and formatted forms). We proceed with providing stratified accuracy results. 

\subsubsection{Accuracy by LLM and task type}
Tables \ref{tab:GPT54Accuracy}, \ref{tab:Gemini3Accuracy}, and \ref{tab:Qwen36Accuracy} present our stratified accuracy results per LLM. Note that the task type ``Composition'' comprises the tasks that handle the composition operators used in HMSCs. We further analyse these results to answer the research questions in the remainder of this section. (There are other stratified accuracy results in the replication package, e.g., based on the subject system, not presented here for the sake of brevity.)

\begin{table}
\caption{Summary of performance per task type for {GPT-5.4-Think Deeper}}\label{tab:GPT54Accuracy}
\begin{center}
\begin{tabular}{lrrrr}
\toprule
Task type & Correct & Incorrect & Total & Accuracy \\
\midrule
Identifying events & 3 & 1 & 4 & 75.0\% \\
Identifying event orders & 5 & 2 & 7 & 71.4\% \\
Abstraction & 2 & 8 & 10 & 20.0\% \\
Traces & 3 & 1 & 4 & 75.0\% \\
LTS & 2 & 2 & 4 & 50.0\% \\
Composition & 5 & 9 & 14 & 35.7\% \\
Overall & 20 & 23 & 43 & 46.5\% \\
\bottomrule
\end{tabular}
\end{center}
\end{table}

\begin{table}
\caption{Summary of performance per task type for {Gemini 3 - Thinking}}\label{tab:Gemini3Accuracy}
\begin{center}
\begin{tabular}{lrrrr}
\toprule
Task type & Correct & Incorrect & Total & Accuracy \\
\midrule
Identifying events & 4 & 0 & 4 & 100.0\% \\
Identifying event orders & 7 & 0 & 7 & 100.0\% \\
Abstraction & 1 & 9 & 10 & 10.0\% \\
Traces & 2 & 2 & 4 & 50.0\% \\
LTS & 0 & 4 & 4 & 0.0\% \\
Composition & 10 & 4 & 14 & 71.4\% \\
Overall & 24 & 19 & 43 & 55.8\% \\
\bottomrule
\end{tabular}
\end{center}
\end{table}

\begin{table}
\caption{Summary of performance per task type for {Qwen 3.6 Plus - Thinking}}\label{tab:Qwen36Accuracy}
\begin{center}
\begin{tabular}{lrrrr}
\toprule
Task type & Correct & Incorrect & Total & Accuracy \\
\midrule
Identifying events & 4 & 0 & 4 & 100.0\% \\
Identifying event orders & 6 & 1 & 7 & 85.7\% \\
Abstraction & 3 & 7 & 10 & 30.0\% \\
Traces & 2 & 2 & 4 & 50.0\% \\
LTS & 1 & 3 & 4 & 25.0\% \\
Composition & 5 & 9 & 14 & 35.7\% \\
Overall & 21 & 22 & 43 & 48.8\% \\
\bottomrule
\end{tabular}
\end{center}
\end{table}

\iffalse

\begin{table} 
\caption{Summary of total accuracy per task type across the three LLMs.}\label{tab:overallTaskAccuracy} 

\begin{tabular}{lrrrrrr}
\toprule
Task type/Accuracy & GPT & Gemini & Qwen \\
\midrule
Identifying events & 75.0\% & 100.0\% & 100.0\% \\
Identifying event orders & 71.4\% & 100.0\% & 85.7\% \\
Abstraction &  20.0\% & 10.0\% & 30.0\% \\
Traces & 75.0\% & 50.0\% & 50.0\% \\
LTS & 50.0\% & 0.0\% & 25.0\% \\
Composition & 35.7\% & 71.4\% & 35.7\% \\
\bottomrule
\end{tabular}
\end{table} 

\fi

\subsection{RQ1: To what extent can LLMs identify and reason about the basic semantic elements of MSCs?}

LLMs performed strongly on the basic semantic concepts of MSCs, in the task required identifying events and event ordering. Across all models, the combined accuracy for these basic semantic tasks was 87.9\% with little variation. Model size does not seem to be an issue for the observed results. Most of the mistakes concerned counting extra events from the environment or incorrectly adding ordering across instances (due to the vertical positioning of events), which may be an artifact of visual analysis. 

\begin{table}[H]
\centering
\caption{Performance on basic MSC semantic tasks.}
\begin{tabular}{lrrr}
\toprule
\textbf{Basic MSC task} & \textbf{Correct} & \textbf{Total} & \textbf{Accuracy} \\
\midrule
Identifying events & 11 & 12 & 91.7\% \\
Identifying event orders & 18 & 21 & 85.7\% \\
\midrule
\textbf{Combined basic semantic tasks} & \textbf{29} & \textbf{33} & \textbf{87.9\%} \\
\bottomrule
\end{tabular}
\end{table}

\rqanswer{LLMs are generally competent at understanding basic MSC semantic concepts. Basic event recognition and event ordering are the strongest areas in the results.}

\subsection{RQ2: How well do LLMs perform semantic transformations over MSCs? }

Semantic transformations were substantially harder than basic semantic recognition. The two transformation-oriented categories were abstraction and composition. Across all models, the combined accuracy for transformation tasks was only 36.1\%.

\begin{table}[H]
\centering
\caption{Performance on MSC semantic transformation tasks.}
\begin{tabular}{lrrr}
\toprule
\textbf{Transformation task} & \textbf{Correct} & \textbf{Total} & \textbf{Accuracy} \\
\midrule
Abstraction & 6 & 30 & 20.0\% \\
Composition & 20 & 42 & 47.6\% \\
\midrule
\textbf{Combined transformation tasks} & \textbf{26} & \textbf{72} & \textbf{36.1\%} \\
\bottomrule
\end{tabular}
\end{table}

Abstraction was the weakest category overall. All three models struggled with it. Gemini achieved only 10.0\%, GPT-5.4-Think Deeper achieved 20.0\%, and Qwen achieved 30.0\%. The main reasons for the observed mistakes were the following: 
\begin{enumerate} 
\item In abstracting from message, LLMs failed to respect the old causal order after abstraction. 
\item When combining instances, they often added arbitrary new orders by ordering events that happened in the two merged instances. 
\item In horizontal composition, they ordered events of the same instance that happened in parallel. LLMs do not seem to have a sound notion of co-regions in MSCs. 
\item In vertical composition, they ordered events across different instances because of their vertical positioning. 

\end{enumerate}

\begin{table}[H]
\centering
\caption{Model-level performance on abstraction and composition.}
\begin{tabular}{lrr}
\toprule
\textbf{Model} & \textbf{Abstraction accuracy} & \textbf{Composition accuracy} \\
\midrule
GPT-5.4-Think Deeper & 20.0\% & 35.7\% \\
Gemini 3 - Thinking & 10.0\% & 71.4\% \\
Qwen 3.6 Plus - Thinking & 30.0\% & 35.7\% \\
\bottomrule
\end{tabular}
\end{table}

%Composition showed more model-dependent variation. Gemini performed relatively well at 71.4\%, whereas GPT-5.4-Think Deeper and Qwen both achieved only 35.7\%. This indicates that composition is not uniformly beyond LLM capability, but model-specific reasoning behaviour matters greatly. Abstraction, by contrast, appears to be a systematic weakness across all models, likely because it requires preserving semantic equivalence while removing or hiding details.

\rqanswer{LLMs perform poorly on MSC semantic transformations overall. Abstraction is a severe weakness, while composition is possible for some models but inconsistent.}

\subsection{RQ3: Can LLMs reason about the behavioural semantics of MSCs?}

LLMs performed poorly in reasoning about behavioural semantics with an overall accuracy of 41.7\%. Calculating all linear orders arising from a partial order and constructing diamond-structures for interleaving events proved challenging. When successful, these are often achieved through generating and executing Python code that encodes the basic semantic concepts and the task involved. Also, the visual structure of LTSs seems to be a problem for LLMs. 

\begin{table}[H]
\centering
\caption{Performance on behavioural semantics tasks.}
\begin{tabular}{lrrr}
\toprule
\textbf{Behavioural task} & \textbf{Correct} & \textbf{Total} & \textbf{Accuracy} \\
\midrule
Traces & 7 & 12 & 58.3\% \\
LTS & 3 & 12 & 25.0\% \\
\midrule
\textbf{Combined behavioural semantics tasks} & \textbf{10} & \textbf{24} & \textbf{41.7\%} \\
\bottomrule
\end{tabular}
\end{table}

The models showed moderate ability on traces but poor performance on LTS reasoning.

\begin{table}[H]
\centering
\caption{Model-level performance on traces and LTS.}
\begin{tabular}{lrr}
\toprule
\textbf{Model} & \textbf{Traces accuracy} & \textbf{LTS accuracy} \\
\midrule
GPT-5.4-Think Deeper & 75.0\% & 50.0\% \\
Gemini 3 - Thinking & 50.0\% & 0.0\% \\
Qwen 3.6 Plus - Thinking & 50.0\% & 25.0\% \\
\bottomrule
\end{tabular}
\end{table}

%Trace reasoning appears easier than LTS construction or interpretation. This is plausible because traces can often be checked as linear sequences, whereas LTS reasoning requires representing branching behaviour, states, transitions, and possibly nondeterminism.

\rqanswer{LLMs can reason about traces to a limited or moderate extent, but they struggle substantially with LTS-based behavioural semantics. }

\subsection{RQ4: How does the performance vary across LLMs model complexity and task complexity?}

Overall model performance was fairly close across LLMs, with Gemini achieving the highest aggregate accuracy.  The results do not show a simple monotonic relationship in which the apparently more advanced or deeper-thinking model is always better. 
%Gemini had the best overall score, but its performance was uneven: it was excellent on event identification, event ordering, and composition, but failed all LTS tasks. GPT-5.4-Think Deeper was weaker overall, but strongest on traces and better than the others on LTS. 
The differences among LLMs can probably be attributed to better code generation capabilities to express the higher-level semantic reasoning tasks into Python code exploiting the basic semantic concepts. The overall accuracies are provided in Table \ref{tab:overallAccuracyRQ4}. Task complexity had a more pronounced effect on performance, as observed before. Model size does not seem to influence performance significantly.

\begin{table}[H]
\centering
\caption{Overall model performance.}\label{tab:overallAccuracyRQ4}
\begin{tabular}{lrrr}
\toprule
\textbf{Model} & \textbf{Correct} & \textbf{Total} & \textbf{Overall accuracy} \\
\midrule
Gemini 3 - Thinking & 24 & 43 & 55.8\% \\
Qwen 3.6 Plus - Thinking & 21 & 43 & 48.8\% \\
GPT-5.4-Think Deeper & 20 & 43 & 46.5\% \\
\bottomrule
\end{tabular}
\end{table}

\iffalse 

\begin{table}[H]
\centering
\caption{Accuracy by broad task-complexity group.}
\begin{tabular}{lr}
\toprule
\textbf{Task group} & \textbf{Accuracy} \\
\midrule
Basic semantic tasks: events and event orders & 87.9\% \\
Behavioural semantics: traces and LTS & 41.7\% \\
Transformations: abstraction and composition & 36.1\% \\
\bottomrule
\end{tabular}
\end{table}
\fi

\rqanswer{Different LLMs show similar performance. Task complexity explains performance better than model size. }

\subsection{RQ5: Which MSC concepts constitute recurring pain patterns for LLMs?}

The clearest recurring pain patterns are tasks that involve respecting event orders across model transformations. This pattern manifests itself in abstraction, composition, trace-, and LTS generation and accounts for a majority of the observed mistakes.  
This can be further attributed to lack of a sound model of co-regions and explicit causal relations, as well as limited internal reasoning capabilities. The latter is typically mitigated by generating Python code that performs the reasoning, but then sometimes the LLM produced incorrect code not capturing the correct reasoning process. Connecting LLMs to sound reasoning tools for such tasks and providing a feedback loop should address this pain point.  

\iffalse

\begin{table}[H]
\centering
\caption{Recurring MSC pain patterns.}
\begin{tabularx}{\textwidth}{l r X}
\toprule
\textbf{Pain pattern} & \textbf{Accuracy} & \textbf{Interpretation} \\
\midrule
Abstraction & 20.0\% & Severe recurring failure across all models. \\
LTS & 25.0\% & Severe difficulty with state-transition behavioural semantics. \\
Composition & 47.6\% & Borderline but still mostly incorrect overall. \\
Complex composition examples, especially Example 8 & 0.0\% & Failed by all models. \\
\bottomrule
\end{tabularx}
\end{table}

The strongest pain pattern is abstraction. Every model failed most abstraction tasks: Gemini scored 10.0\%, GPT-5.4-Think Deeper scored 20.0\%, and Qwen scored 30.0\%. This suggests that abstraction is not just a model-specific weakness but a general limitation.

The second major pain pattern is LTS reasoning. Gemini failed all LTS tasks, Qwen solved only one out of four, and GPT-5.4-Think Deeper solved two out of four. This indicates that LLMs often struggle to convert MSC semantics into an explicit behavioural model.

The third pain pattern is composition, especially in more complex examples. Although Gemini performed well on composition overall, GPT-5.4-Think Deeper and Qwen did not. Example 8 was failed by all models, making it a strong candidate for a structurally difficult MSC composition case.
\fi

\rqanswer{The recurring MSC pain patterns concern respecting partial orders across different model transformations. Lack of sound semantic concepts about partial orders in concurrency and limited reasoning capabilities are likely to be the main causes. }

\section{Reflections}\label{sec:reflection}

\subsection{Threats to Validity}
\label{subsec:threats-to-validity}

Several threats to validity should be considered. First, the use of lecture-note examples provides a reliable source of well-formed models and reference solutions, but it also introduces the possibility that some models may have encountered unofficial copies of the material during training. Second, the benchmark contains nine subject models, which is sufficient for an exploratory study but should be expanded in future work to support stronger statistical conclusions. Third, correctness assessment may involve semantic judgement, particularly for diagram transformations and labelled transition systems; therefore, future versions of the benchmark should aim to define more precise and mechanised checking procedures. Also the evaluation focuses on a selected set of LLMs, and results may change as new models and reasoning interfaces become available. Finally, it is unclear how many of the observed mistakes are due to the visual nature of the models and whether resorting to textual representations will reveal the reasoning limitations more clearly. 

\subsection{Environmental Sustainability} 
The amount of CO2 emitted while performing this research has been a major unaddressed concern. 
Our past research \cite{CheungKJ0Z25} demonstrated that LLMs are likely to have an order of magnitude more of CO2 emission compared to manual for comparable tasks compared to manual development. We did attempt to analyse how much CO2 emission the experiments and the LLM assistance in the analysis of the paper results have cost, but the present tools are very limited, particularly in estimating the amount of CO2 emission arising from prior training (and amortising it on inference queries). 
This, as well as other environmental and economic sustainability impacts, remains a major concern for our future research. 

\subsection{Declearation of the Use of AI} 
We have used AI (GPT 5.5 - Thinking) to formulate the research questions, validate the consistency of queries and judgments across all experiments, analyse and summarise the judgments, and summarise the answers to the research questions.  The prompts used for the experiments and their validations are publicly available in the replication package: \url{https://smrmousavi.github.io/sjouke-fest/}. The formatting and the landing page of the replication package were also generated using AI (GPT 5.5 - Thinking). The introduction, related work, basic (semantic) concepts,  and formulation of tasks and prompts were done manually by the author. Tables \ref{tab:code-semantics-comparison} and \ref{tab:architecture-understanding-comparison}   summarising the related work were developed using AI (GPT 5.5 - Thinking)  and edited and summarised afterwards. The author has thoroughly reviewed and edited the AI-generated content mentioned above and takes responsibility for the entire content of the paper.

\subsection{Future Plans}
Our current results suggest that LLMs can handle basic MSC concepts, but struggle with abstraction, composition, concurrency, and interleaving. These findings motivate future work on formulating and testing concrete hypotheses about recurring LLM ``pain patterns'' in formal diagram understanding.

\subsection*{Acknowledgments}
Mohammad Reza Mousavi has been
partially supported by ITEA/InnovateUK projects GENIUS and GreenCode.


 \bibliographystyle{splncs04}
 \bibliography{hmsc}

\end{document}